# Magnetism and pairing in Hubbard bilayers

Raimundo R. dos Santos[*]
Instituto de Física, Universidade Federal Fluminense, 24020-150 Niterói, Rio de Janeiro, Brazil
(February 10, 1995)

We study the Hubbard model on a bilayer with repulsive on-site interactions, $U$, in which fermions undergo both intra-plane ($t$) and inter-plane ($t_z$) hopping. This situation is what one would expect in high-temperature superconductors such as YBCO, with two adjacent $CuO_2$ planes. Magnetic and pairing properties of the system are investigated through Quantum Monte Carlo simulations for both half- and quarter-filled bands. We find that in all cases inter-planar pairing with $d_{x^2-z^2}$ symmetry is dominant over planar pairing with $d_{x^2-y^2}$ symmetry, and that for $t_z$ large enough pair formation is possible through antiferromagnetic correlations. However, another mechanism is needed to make these pairs condense into a superconducting state at lower temperatures. We identify the temperature for pair formation with the spin gap crossover temperature.

71.27.+a, 74.25.Dw, 74.20.Mn, 75.10.Jm, 75.10.Lp

## I. INTRODUCTION

The normal state of the high temperature superconductors unveils interesting features, especially in the YBCO compounds. Firstly, the in-plane resistivity of optimally doped (i.e., hole fraction $x = x_m$, corresponding to a maximum $T_c$) samples of $YBa_2Cu_3O_{7-y}$ with $y = 0.10$, displays linear behavior in a wide range of temperatures, from just above $T_c \simeq 90$ K to nearly 1000 K;[1,2] this is to be contrasted with the $T^2$ behavior at low temperatures, that would be expected for a usual Fermi liquid. For underdoped samples (i.e., $x < x_m$), the linear temperature behavior of the resistivity crosses over to $T^\alpha$, with $\alpha \simeq 2.5$, below a characteristic temperature $T^\times(x)$.[3] Overdoped samples, on the other hand, behave like usual metals, in the sense that the resistivity crosses over from a linear-$T$ dependence (at high temperatures) to Fermi liquid behavior at lower temperatures.[4,5] Secondly, the magnetic response – probed by nuclear relaxation rate ($1/T_1T$) on the Cu(2) sites[6,7] and by NMR Knight shift at the $^{89}Y$ sites[8] – also shows different behavior in the underdoped and overdoped regimes. For overdoped samples, the relaxation rate, which is proportional to the *staggered* susceptibility ($\chi(\pi)$), increases monotonically as the temperature is lowered, in a Curie-like fashion;[6,7] this is indicative of gapless spin excitations from a state with strong antiferromagnetic correlations. At $T_c$ one expects the simultaneous formation and condensation of pairs, leading, respectively, to spin and superconducting gaps. The situation is similar for the NMR Knight shift, which is proportional to the *uniform* susceptibility ($\chi(0)$): one observes a Pauli-like behavior, in the sense that it is roughly temperature independent.[3] For underdoped samples both $\chi(\pi)$ and $\chi(0)$ show a marked deviation from their respective Curie and Pauli behaviors as the temperature is lowered; i.e., they start to decrease as $T$ decreases below certain temperature scales $T_\pi(x)$ and $T_0(x)$. While one cannot be precise to the point of identifying all temperature scales, the general expectation is that $T^\times \simeq T_\pi \simeq T_0$. Thus, one may interpret the behavior in the underdoped region as due to the formation of pairs (without condensation) at a temperature $T^\times$, which is accompanied by the opening of a spin gap.[9] As the temperature is lowered further, a superconducting gap opens at $T_c$, giving rise to a superconducting instability. Further, Ito et al.[3] have stressed that the temperature dependence of their in-plane resistivity data, $\rho(T)$, is such that $\chi(0) \sim \rho(T)/T$ and $\chi(\pi) \sim \rho(T)/T^2$. Similar conclusions may be drawn from the analysis of transport properties of $YBa_2Cu_4O_8$, corresponding to the underdoped regime.[10] The interpretation of these data is still a matter of debate. On the one hand, the above mentioned transport (charge) properties of the underdoped materials are strongly influenced by the opening of a spin gap, suggesting inseparability of spin and charge behaviors.[11] On the other hand, one could also assume spin and charge separation in the sense that the spin gap opening would indicate formation of pairs, which would Bose condense (superconduct) at a lower temperature;[9,12] the overdoped regime would then be described as an ordinary Fermi liquid, with both spin and superconducting gaps opening at the same critical temperature, $T_c$.

In order to gain insight into these issues, one works with simplified models which may highlight the main physical mechanisms at work. For instance, one possibility is to treat the planar spin excitations as those of a nearly antiferromagnetic Fermi liquid (NAFL).[13,14] This approach has been used to discuss several properties of *overdoped* compounds,[13–16] though it is not clear at the moment how to incorporate the spin gap in the treatment of the *underdoped* regime. An alternative explanation for the experimental data is based on the presence of a double $CuO_2$ layer in YBCO compounds, as opposed to the single layered structure of the $La_{2-x}Sr_xCuO_4$ materials.[17] While the spin gap behavior in underdoped La compounds was attributed[17] to a spin-density wave (SDW) instability, it was suggested that antiferromagnetic correlations between fermions in adjacent planes in bilayer materials lead to singlet pairing with each member of the pair lying on each plane; this, in turn, would be responsible for the non-Fermi liquid behavior. These ideas[17–19] have been tested on models (e.g., the $t$-$J$ model) with antiferromagnetic Heisenberg-like coupling between spins on different layers,[12,17,20,21] and they indeed lead to pairing between fermions in different planes. Unfortunately, interplane couplings of the order of (and, in some cases, even larger than) intraplane couplings are generally needed to achieve pairing, which seems rather unrealistic. The Hubbard model on a bilayer, with on-site repulsion and both intra- and inter-plane hopping, can be considered as a weaker-coupling version of the models mentioned above, and should therefore provide a more realistic description of the actual physical situation. The Hamiltonian is

$$\mathcal{H} = -\sum_{\langle i,j \rangle \atop \sigma} t_{ij} \left( c_{i\sigma}^\dagger c_{j\sigma} + \text{H.c.} \right)$$
$$+ U \sum_i \left( n_{i\uparrow} - \frac{1}{2} \right) \left( n_{i\downarrow} - \frac{1}{2} \right) - \mu \sum_{i,\sigma} n_{i\sigma} , \quad (1)$$

where the sums run over sites of two square lattice layers, $\langle i,j \rangle$ denotes nearest neighbor sites, and the hopping integral is given by

$$t_{ij} = \begin{cases} t & \text{if } i \text{ and } j \text{ are within the same plane}, \\ t_z & \text{if } i \text{ and } j \text{ lie in different planes}. \end{cases} \quad (2)$$

$c_{i\sigma}^\dagger$ ($c_{i\sigma}$) creates (annihilates) a fermion at site $i$ with spin $\sigma$, H.c. stands for Hermitian conjugate, $U > 0$ is the on-site repulsion, and $\mu$ is the chemical potential controlling the band-filling, $\langle n \rangle$. This model has been studied previously by Bulut et al.[22], who found evidence for nodeless singlet pairing, from random-phase-approximation–Eliashberg calculations based on the exchange of spin fluctuations, for $\langle n \rangle = 1$ and 0.85; their results for half-filling were compared with those from Monte Carlo simulations on a 2×(4×4) lattice. In view of the important role interlayer pairing may play as a mechanism for high-temperature superconductivity, independent checks should be made in order to assure the effect is independent of model details and of the approximations used. With this in mind, here we report the results of extensive Monte Carlo simulations, in which both magnetic and pairing properties of the model are examined for larger lattices [up to 2×(8×8)] and for the cases of half- and quarter-filled bands.

The layout of the paper is as follows. In Sec. II we mention the difficulties posed by the "minus sign problem" and present the results for magnetic susceptibilities and correlation functions. Similarly, Sec. III deals with the superconducting susceptibilities and correlation functions. Finally, Sec. IV summarizes our findings and presents the conclusions.

## II. MAGNETIC PROPERTIES

In a grand-canonical quantum Monte Carlo simulation[23–27] the imaginary time is discretized through the introduction of $M$ "time" slices, with $M \propto 1/T$. The "Boltzmann weight" is given by a product of fermion determinants, and is only positive definite for the repulsive model at half-filling and for the attractive model at any filling;[24,27,28] otherwise, some configurations will give negative contribution. This "minus sign problem" is circumvented by redefining the averages in such way that an average sign appears in the denominator.[29] The average sign, $\langle \epsilon \rangle'$, behaves generically as follows:[30] for a fixed temperature, it drops dramatically from 1 very near half-filling, reaching a minimum for some band filling, and eventually increasing to $\langle \epsilon \rangle \simeq 1$ for some occupation near $\langle n \rangle = 1/2$; the value of $\langle n \rangle$ for which this happens is not very sensitive to details such as the values of $U$, of $t_z$, and so on.[30] As one dopes further away, the average sign drops again. Moreover, as the temperature decreases, the dip in $\langle \epsilon \rangle$ becomes more pronounced, reaching a very small value; this introduces uncontrollable noise in the calculation of average values. In view of this, we restrict our discussion to half- and quarter-filled bands (i.e., to $\langle n \rangle = 1$ and 1/2, respectively).

The clusters used here have $N_s = 2 \times (L \times L)$ sites, with periodic boundary conditions (PBC); that is, each site is connected with its six nearest neighbors through a hopping term. Due to the PBC along a direction of finite size $L_z = 2$, the effective hopping along the $z$ direction becomes $t_\perp = 2t_z$. The simulations were performed on Sun and IBM RISC 600/525 workstations, and on a CRAY Y-MP/2E; a single datum point involves between 500 and 2000 MC sweeps over all time slices and we took $\Delta\tau \equiv \beta/M = 0.125$.

Magnetic properties are examined through both the magnetic susceptibility,

$$\chi(\mathbf{q}) = \frac{1}{N_s} \sum_{i,j} e^{i\mathbf{q}\cdot(\mathbf{r}_i - \mathbf{r}_j)} \int_0^\beta d\tau \, \langle m_i(\tau) m_j(0) \rangle , \quad (3)$$

with

$$m_i(\tau) = e^{\tau\mathcal{H}} [n_{i\uparrow} - n_{i\downarrow}] e^{-\tau\mathcal{H}} , \quad (4)$$

and the magnetic structure factor,

$$S(\mathbf{q}) = \frac{1}{N_s} \sum_{i,j} e^{i\mathbf{q}\cdot(\mathbf{r}_i - \mathbf{r}_j)} \langle m_i m_j \rangle , \quad (5)$$

where $m_i \equiv m_i(0)$.

Figure 1 shows the uniform ($\mathbf{q} = \mathbf{0}$) and staggered ($\mathbf{q} = \pi$) susceptibilities for the half-filled band case, with $t_\perp = 0.7$ and $U = 4$ (from now on, energies will be expressed in units of $t$, the nearest-neighbor planar hopping integral, and temperatures in units of $t/k_B$, $k_B$ being the Boltzmann constant); the system sizes are $2 \times (L \times L)$, with $L = 4, 6$ and 8. The uniform susceptibility (see Fig.

1(a)) decreases as the temperature decreases, in a non-Pauli fashion; this should be contrasted with the behavior of the staggered susceptibility (see Fig. 1(b); notice the difference in scales), which shows a Curie-like behavior and scaling with lattice size, $L$. By analogy with the single-layer Hubbard model,[24,31] one associates an insulating antiferromagnetic state also for the bilayer at half filling.

maxima and roughly the same heights, though one may argue that as $t_\perp$ increases, the maximum in the $q_z = 0$ sector narrows slightly, while that in the $q_z = \pi$ sector flattens out also slightly. The important point is that antiferromagnetic correlations are always dominant, at least in one (planar) direction, signalled by $S(\mathbf{q})$ being roughly the same, as long as $q_x = \pi$.

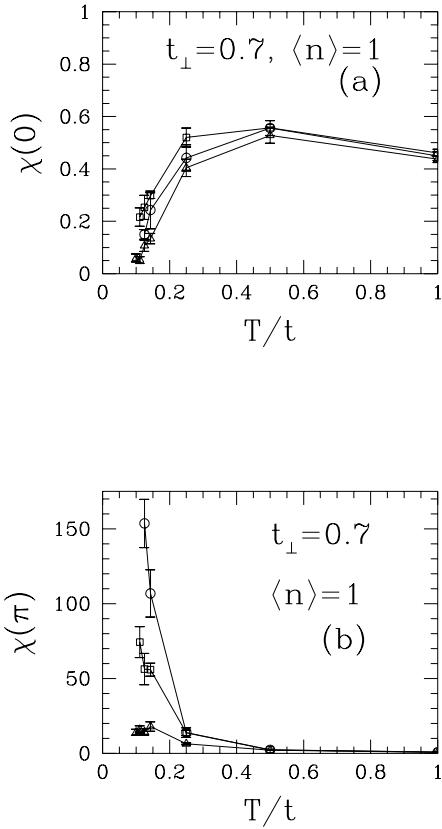

FIG. 1. Uniform (a) and staggered (b) magnetic susceptibilities as functions of temperature, for the repulsive Hubbard bilayer at half-filling, with $U = 4$ and $t_\perp = 0.7$. System sizes are $2 \times (L \times L)$, with $L = 4$ (triangles), $L = 6$ (squares), and $L = 8$ (circles). The lines are guides to the eye.

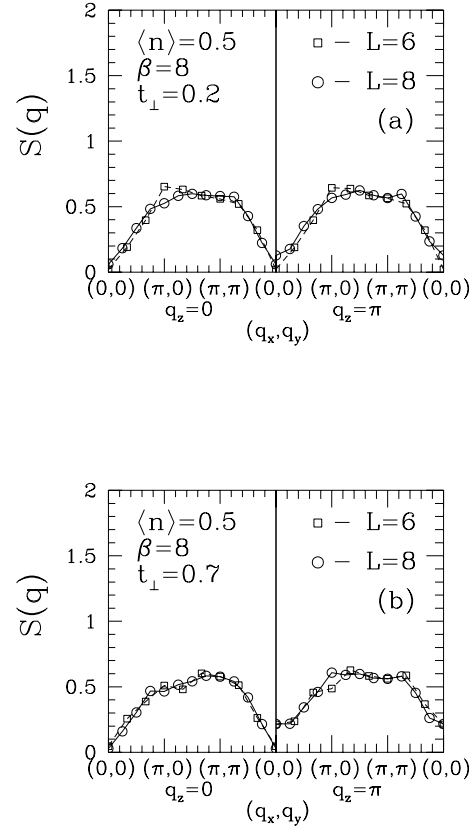

FIG. 2. Wavevector dependence of the structure factor for a fixed inverse temperature, $\beta = 8$, for the repulsive Hubbard bilayer at quarter-filling, with $U = 4$ and $t_\perp = 0.2$ (a) and $t_\perp = 0.7$ (b). System sizes are $2 \times (L \times L)$, with $L = 6$ (squares), and $L = 8$ (circles). The error bars are smaller than the data points, and the lines are guides to the eye.

As mentioned before, the minus-sign problem prevents us from going to very low temperatures away from half filling; the optimum choice for the bilayers i.e., the one that allows us to reach the lowest temperatures with an acceptable $\langle \epsilon \rangle \geq 0.6$, is $\langle n \rangle = 0.5$. In order to discuss the magnetic correlations for this filling, in Fig. 2 we display the structure factor, Eq. (5), at fixed temperature; since the data are symmetric under the exchange $(q_x, q_y) \to (q_y, q_x)$, the Brillouin zone is only shown partially. In each of the $q_z$ sectors the behavior is similar for $t_\perp = 0.2$ [Fig. 2(a)] and for $t_\perp = 0.7$ [Fig. 2(b)], with very broad

It is also instructive to discuss the $\mathbf{q}$-dependence of the susceptibility. For $t_\perp = 0.2$ (see Fig. 3(a)), two peaks with approximately the same height appear in both $q_z = 0$ and $q_z = \pi$ sectors, indicating that the magnetic response is the same whether the field is uniform or staggered along the $z$ direction (perpendicular to the two layers). As one increases the interlayer hopping to $t_\perp = 0.7$ (see Fig. 3(b)), only one peak is found, corresponding to a field staggered in the $z$-direction. These peaks come about as a result of correlations adding up coherently; we shall return to this point below.

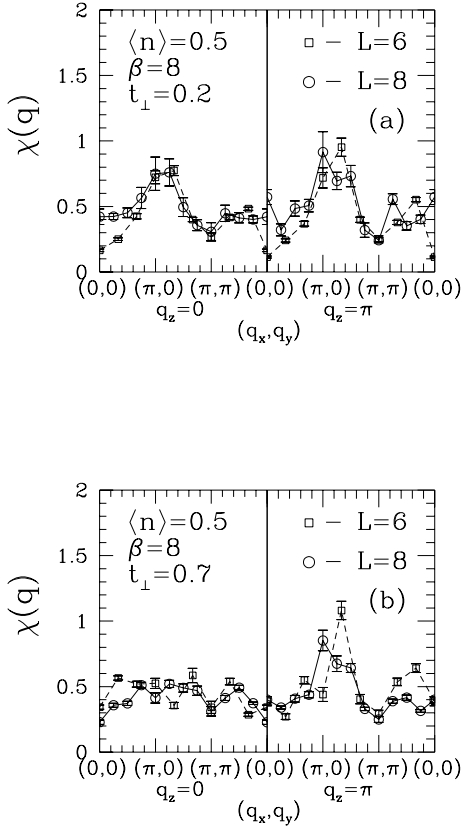

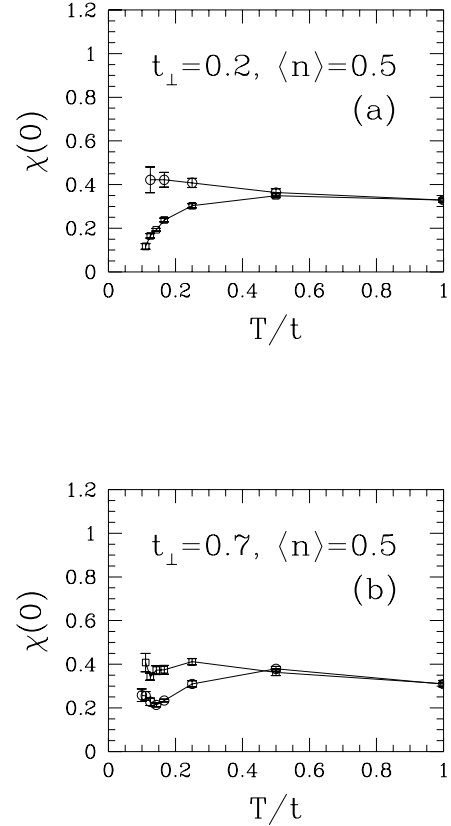

FIG. 3. Wavevector dependence of the susceptibility for a fixed inverse temperature, $\beta = 8$, for the repulsive Hubbard bilayer at quarter-filling, with $U = 4$ and $t_\perp = 0.2$ (a) and $t_\perp = 0.7$ (b). System sizes are $2 \times (L \times L)$, with $L = 6$ (squares), and $L = 8$ (circles). The lines are guides to the eye.

Figure 4(a) shows the temperature dependence of the uniform susceptibility at quarter-filled band for $t_\perp = 0.2$, and for $L = 6$ and 8. From the data for $L = 6$ alone, one might be tempted to infer a spin gap behavior, signalled by the downturn of $\chi(0)$, together with the absence of long range magnetic order. This behavior, however, does not seem to persist for the $L = 8$ system; on the contrary, the weak temperature dependence of $\chi(0)$ suggests a Pauli-like — and therefore metallic — behavior for the whole temperature range examined. The data for $t_\perp = 0.7$, shown in Fig. 4(b), can be interpreted differently. As the temperature decreases, a downturn in $\chi(0)$ is followed by an increase at lower temperatures; unlike what was observed in Fig. 4(a), this feature is independent of system size. While this is not the prototypical spin gap behavior, the different temperature dependences found for the two values of $t_\perp$ can hardly be regarded as fortuitous. This, together with the above analysis of $\chi(\mathbf{q})$, suggests that the nature of spin excitations changes as one increases $t_\perp$. In the following section we discuss the possible bearings of the magnetic properties on pairing.

FIG. 4. Uniform susceptibility as a function of temperature, for the repulsive Hubbard bilayer at quarter-filling, with $U = 4$ and $t_\perp = 0.2$ (a) and $t_\perp = 0.7$ (b). System sizes are $2 \times (L \times L)$, with $L = 6$ (squares), and $L = 8$ (circles). The lines are guides to the eye.

### III. SUPERCONDUCTING PROPERTIES

Superconducting properties are probed by the uniform ($\mathbf{q} = 0$) zero-frequency pair susceptibilities,

$$P_\lambda = \int_0^\beta d\tau \, \langle \Delta_\lambda(\tau) \Delta_\lambda^\dagger(0) \rangle , \quad (6)$$

and by equal-time uniform ($\mathbf{q} = 0$) correlation functions,

$$\mathcal{C}_\lambda \equiv \langle \Delta_\lambda^\dagger \Delta_\lambda + \Delta_\lambda \Delta_\lambda^\dagger \rangle , \quad (7)$$

where the pair-field operator is given by

$$\Delta_\lambda^\dagger = \frac{1}{\sqrt{N_s}} \sum_{\mathbf{k}} f_\lambda(\mathbf{k}) \, c_{\mathbf{k}\uparrow}^\dagger c_{-\mathbf{k}\downarrow}^\dagger , \quad (8)$$

with $f_\lambda(\mathbf{k})$ defining its symmetry. It is by now well established[32-35] that the dominant pairing correlations in the *planar* Hubbard model have $d_{x^2-y^2}$-symmetry, for which

$$f_{d_{x^2-y^2}}(\mathbf{k}) = \cos k_x - \cos k_y \ . \qquad (9)$$

Our strategy here is to compare this dominant planar pairing with inter-planar correlations of different symmetries, such as

$$f_{d_z}(\mathbf{k}) = \cos k_z \ , \qquad (10)$$

$$f_{d_{x^2-z^2}}(\mathbf{k}) = \cos k_x - \cos k_z \ , \qquad (11)$$

$$f_{s_z^*}(\mathbf{k}) = \cos k_x + \cos k_z \ , \qquad (12)$$

$$f_{d_{xz}}(\mathbf{k}) = \sin k_x \sin k_z \ , \qquad (13)$$

as well as with other combinations of longer range.

As it turned out, the dominant susceptibility corresponding to interplanar pairing has $d_{x^2-z^2}$ symmetry; that is, it is larger than any other in all situations (i.e., different temperatures, interlayer hoppings, and band fillings) examined, especially the one corresponding to $d_z$ (the so-called nodeless $d$-wave[22]). We have also checked that the largest planar pairing susceptibility has $d_{x^2-y^2}$ symmetry even in the case of a bilayer. In what follows, we therefore concentrate our discussion in terms of the dominant ones.

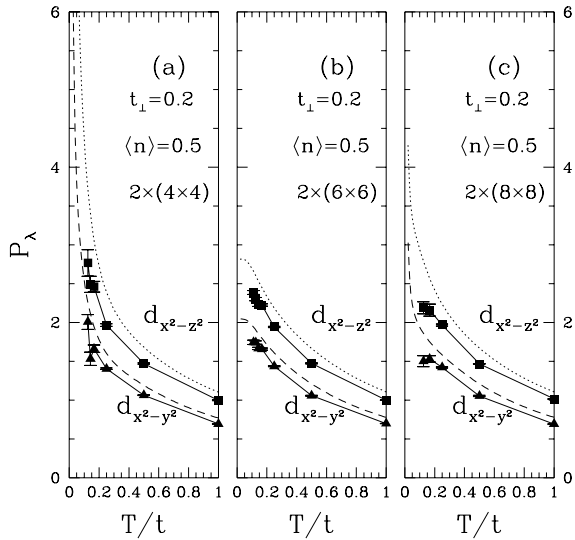

FIG. 5. Temperature dependence of the pairing susceptibilities at quarter-filling, for $t_\perp = 0.2$, for different system sizes: $2 \times (L \times L)$, with (a) $L = 4$, (b) $L = 6$, and (c) $L = 8$. Data for the interacting case with $U = 4$ [non-interacting], are represented by squares [dotted] for $d_{x^2-z^2}$ symmetry, and by triangles [dashed] for $d_{x^2-y^2}$ symmetry; solid lines are guides to the eye.

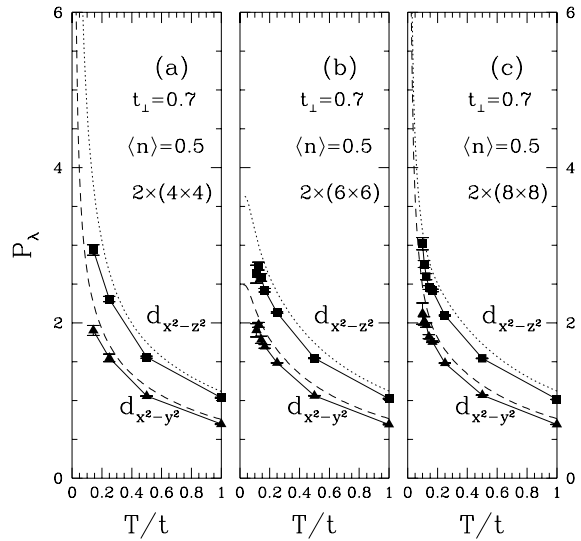

FIG. 6. Same as Fig. 5, but for $t_\perp = 0.7$

Figure 5 shows the temperature dependence of $P_{d_{x^2-y^2}}$ and $P_{d_{x^2-z^2}}$, for the interacting ($U = 4$) and free ($U = 0$) cases, for different system sizes, at quarter filling, and with $t_\perp = 0.2$. In all cases, the susceptibilities are suppressed by the presence of a repulsive on-site interaction; this means that the fermions are less likely to pair in the presence of the on-site repulsion, similarly to the single-layer Hubbard model.[32] Nevertheless, it is important to note that interplanar pairing always dominates the planar one. As one increases the interlayer hopping, an interesting feature emerges. In Fig. 6, for $t_\perp = 0.7$, the susceptibilities in the interacting case grow faster at lower temperatures than those corresponding to the free case, for the largest system examined. One therefore expects that there is a temperature, $T^* \sim 0.1$, below which the interacting susceptibility is larger than the free one, indicating a tendency towards pairing arising from repulsion. In view of the analysis of Sec. II, one concludes that pairing is favored when "time"-correlations along the $z$ direction are predominantly of antiferromagnetic nature; that is, when the degeneracy between uniform and staggered correlations is broken.

In order to verify whether the system could be close to an actual phase transition, we have also examined the dependence of the pairing correlation function, Eq. (7), with the system size. If an infinite two-dimensional system (or bilayer) undergoes a superconducting transition, it belongs to the Kosterlitz-Thouless[36] $XY$-model universality class. Accordingly, pairing correlations should become critical at a critical temperature $T_c$, and decay algebraically,

$$G(r) \sim r^{-\eta} \ , \qquad (14)$$

with $\eta = 0.25$ for $T \to T_c^+$. From finite-size scaling (FSS) theory,[37] one then infers that for a system of linear size

$L$, its associated uniform Fourier transform becomes

$$\mathcal{C}_\lambda \sim \int_0^L d^2r\, r^{-\eta} \sim L^{2-\eta}\,. \qquad (15)$$

Above criticality, the apropriate scaling variable[37] is $L/\xi$, where $\xi \sim \exp(A/\sqrt{T-T_c})$, with $A$ being of order unity, is the correlation length for the infinite system. Therefore, one can assume the following FSS ansatz:

$$\mathcal{C}_\lambda(T,L) = L^{2-\eta} F(L/\xi)\,, \qquad (16)$$

where $F(z)$ is a scaling function such that $F(z) \to$ constant when $L \ll \xi$, recovering Eq. (15). At $T_c$, $\xi = \infty$, so that $L^{\eta-2}\mathcal{C}_\lambda(T_c,L)$ is a constant independent of lattice size. By plotting $L^{\eta-2}\mathcal{C}_\lambda(T,L)$ as a function of $T$ for systems of different sizes, a phase transition would be signalled by a crossing of curves corresponding to systems of different size.[38] Figure 7 shows the size-scaled pairing correlation function with $d_{x^2-z^2}$ symmetry, as a function of the inverse temperature. For all sizes studied $L^{-7/4}\mathcal{C}$ stabilizes to a constant value at large $\beta$, without any indication of crossing. The possibility of a phase transition is therefore ruled out, and the behavior of the pairing susceptibility should be attributed to *pair formation*. These preformed pairs should then condense — through a mechanism absent in the present model — at a lower temperature into a superconducting state.

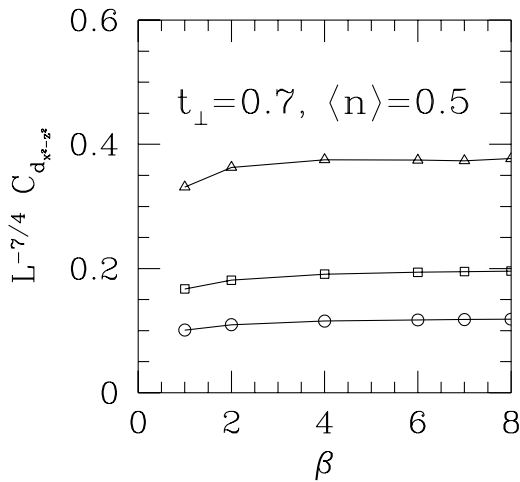

FIG. 7. Inverse-temperature dependence of the size-scaled uniform $d_{x^2-z^2}$-pairing correlation function at quarter-filling, for $t_\perp = 0.7$ and different system sizes: $2\times(L\times L)$, with $L = 4$ (triangles), $L = 6$ (squares), and $L = 8$ (circles). Solid lines are guides to the eye.

## IV. CONCLUSIONS

We have examined the interplay between magnetism and pairing in the Hubbard model on a bilayer. We have found that, even though the magnetic structure factor is not sharply peaked at any particular wavevector, some correlations add up coherently, giving rise to peaks in the **q**-dependent susceptibility. For smaller interlayer hopping, the system's response is similar whether the field is staggered or uniform along the direction perpendicular ($z$) to the bilayer. For larger interlayer hopping this degeneracy is broken, and the system is only distinctively responsive to a $z$-staggered field.

With respect to pairing, we have found that interlayer correlations with $d_{x^2-z^2}$ symmetry are the dominant ones, in the sense that their associated susceptibilty is larger than any other, including planar ones. Thus, the suggestion by Milllis and Monien[17] has been confirmed away from the strong-coupling regime. We have also found evidence that larger interlayer hopping (i.e., $t_\perp = 0.7$) may lead to pairing at low temperatures, and associate this to the dominant antiferromagnetic ("time"-) correlations between the two layers. On the other hand, a finite-size scaling analysis of the pairing correlations indicates the absence of a phase transition into a state with long range (or quasi-long range) order. Upon completion of this work, we have become aware of two recent studies. Hetzel et al.[39] discussed Hubbard bilayers at zero temperature, and have also found no evidence for off-diagonal long range order in the model. Scalettar et al.[40] have considered the half-filled case at finite temperatures, including different chemical potentials on each plane, and have also ruled out off-diagonal long-range order.

The picture that emerges is that pair formation in Hubbard bilayers is possible through antiferromagnetic correlations, though another mechanism is needed to make these pairs condense into a superconducting state at lower temperatures. In this respect, the situation is similar to what happens in the underdoped regime of YBCO compounds, as mentioned in the Introduction. The temperature $T^*$, at which the interacting susceptibility becomes larger than the non-interacting one should be associated with $T^\times$, the spin-gap crossover temperature. Indeed, if one takes $T^* = 0.05t$ and typical values for the Hubbard model parameters ($t \sim 0.5$ eV, $U \simeq 4t \sim 2$ eV), we obtain the estimate $T^* \sim 250$ K, which is of the correct order of magnitude for $T^\times$ in YBCO. Finally, we should mention that this picture has appealing similarities with the results obtained from the *attractive* Hubbard model; see e.g., Ref. 41.

## ACKNOWLEDGMENTS

The author is grateful to the Centro Nacional de Supercomputação of the Federal University of Rio Grande do Sul for a substantial amount of Cray time, and to the Physics Department and RioDatacentro at PUC/Rio, for the use of their computational facilities. Financial support from the Brazilian Agencies, Financiadora de Estudos e Projetos (FINEP), Conselho Nacional de Desen-volvimento Científico e Tecnológico (CNPq) and Coordenação de Aperfeiçoamento do Pessoal de Ensino Superior (CAPES), is also gratefully acknowledged.